\documentclass[twocolumn,showpacs, superscriptaddress]{revtex4} 
\usepackage{graphicx}
\usepackage{amssymb} 
\usepackage{amsmath}
\usepackage{epstopdf}
\usepackage{color}

\begin{document} 

\title{Effect of surface recombination on electroluminescence and photoconversion in a-Si:H/c-Si heterojunction solar cells}

\author{A.V.~Sachenko}
\affiliation{V. Lashkaryov Institute of Semiconductor Physics, NAS of Ukraine, 03028 Kiev, Ukraine}
\author{A.V.~Bobyl}
\affiliation{A.F. Ioffe Physical-Technical Institute RAS, 194021 St.-Petersburg, Russia}
\author{V.N.~Verbitskiy}
\affiliation{A.F. Ioffe Physical-Technical Institute RAS, 194021 St.-Petersburg, Russia}
\author{V.M.~Vlasyuk}
\affiliation{V. Lashkaryov Institute of Semiconductor Physics, NAS of Ukraine, 03028 Kiev, Ukraine}
\author{D.M.~Zhigunov}
\affiliation{M. Lomonosov Moscow State University, 119991 Moscow, Russia}
\author{V.P.~Kostylyov}
\affiliation{V. Lashkaryov Institute of Semiconductor Physics, NAS of Ukraine, 03028 Kiev, Ukraine}
\author{I.O.~Sokolovskyi}
\affiliation{V. Lashkaryov Institute of Semiconductor Physics, NAS of Ukraine, 03028 Kiev, Ukraine}
\affiliation{Department of Physics and Physical Oceanography, Memorial University of Newfoundland, St. John's, NL, A1B 3X7 Canada}
\author{E.I.~Terukov}
\affiliation{A.F. Ioffe Physical-Technical Institute RAS, 194021 St.-Petersburg, Russia}
\affiliation{M. Lomonosov Moscow State University, 119991 Moscow, Russia}
\author{P.A.~Forsh}
\affiliation{M. Lomonosov Moscow State University, 119991 Moscow, Russia}
\author{M.\,Evstigneev}
\email[E-mail: ]{mevstigneev@mun.ca}
\affiliation{Department of Physics and Physical Oceanography, Memorial University of Newfoundland, St. John's, NL, A1B 3X7 Canada}

\begin{abstract}
Surface recombination affects both light-to-electricity and electricity-to-light conversion in solar cells (SCs). Therefore, quantitative analysis and reduction of surface recombination is an important direction in SC research. In this work, electroluminescence (EL) intensity and photoconversion efficiency of a set of 93 large-area (239\,cm$^2$) a-Si:H/c-Si heterojunction SCs (HJSCs) are measured under AM1.5 conditions at 298 K. The HJSC samples differed only in surface recombination velocity, $S$, but otherwise were identical. Variation in $S$ was due to the variation of the chemical conditions under which the samples were treated. It is established that EL quantum efficiency, is affected by $S$ much more strongly than photoconversion efficiency, $\eta$: namely, the reduction of the latter from 20.5\% to 18\% due to an increase of $S$ is accompanied by a decrease of the former by more than an order of magnitude. In HJSCs with well passivated surfaces, i.e. low $S$, EL efficiency reached 2.1\%, which is notably higher than the known values in silicon homojunction diodes. For temperature-dependent measurements of EL and dark I-V curves, one of the samples was cut into small-area (1 cm$^2$) pieces. It was found that EL intensity as a function of temperature develops a maximum at $T$ = 223 K. At low temperatures, the current at weak bias is shown to be due to tunneling mechanism. A theoretical model is developed that explains all these findings quantitatively.
\end{abstract}

\maketitle

\section{Introduction}
Among its numerous applications, electroluminescence (EL) can be used as an efficient tool to investigate the quality of silicon solar cells (SCs) \cite{Kirchartz16, Rau07, Raguse15}, in particular, their bulk recombination parameters \cite{Kirchartz08, Fuyuki05, Wurfel07}. Another major factor that limits SC output power is surface recombination; hence, surface passivation and quantitative characterization of surface recombination velocity, $S$, is an important area of activity in SC research \cite{Hayakawa11, Ogita11, Koyama11, Schuttauf11, Rajesh12}. Yet, there are not many publications investigating the influence of surface recombination on luminescence in SCs. 

We note that this research is of practical importance not only in the field of solar energy, but also in the area of semiconductor light-emitting diodes. Indeed, it has been established in several publications on edge luminescence in silicon diode structures that their photoluminescence quantum efficiency at room temperatures can be as high as about 10\% \cite{Trupke03}, and EL quantum efficiency can reach 1\% \cite{Green01, Ng01}. These high values are surprising in view of the fact that silicon is an indirect-bandgap semiconductor with relatively low radiative recombination constant. In order to achieve these results, efficient surface passivation was crucial \cite{Trupke03}. 

We are aware of only one paper \cite{Takahashi07} where the effect of surface recombination rate on EL intensity was studied systematically. An interesting conclusion made in that work is that EL intensity is very sensitive to surface recombination velocity, and therefore can be used to determine the latter.

The development of heterojunction solar cells (HJSCs) based on a-Si:H/c-Si structures with record efficiency $\eta \ge 20\,\%$ \cite{Jano13} motivates investigation of the interrelation between EL intensity, photoconversion efficiency, and surface recombination velocities in these devices as well. Here, we explore this interrelation both theoretically and experimentally.

On the experimental side, a series of ninety three HJSC samples, identical in all respects except for surface recombination velocity, $S$, was prepared. Variation of $S$ among the samples was due to the variation of the chemical conditions under which the surfaces of our SCs were treated. We measure EL intensity as a function of applied bias in these samples, as well as their spectral characteristics, dark current-voltage curves, and photoconversion efficiency. Also, we develop a theoretical model that can be applied to quantitatively describe both photoconversion and EL within a single mathematical formalism. 

We find, in agreement with \cite{Takahashi07}, that surface recombination has a strong effect on EL. At the same time, photoconversion efficiency depends on $S$ much weaker, at least within the practically relevant range of $S$ less than 200\,cm/s. We show both experimentally and theoretically that EL as a function of temperature develops a maximum due to the competition between the temperature dependence of the radiative recombination coefficient at high temperatures and electron freeze-out at low temperatures. The experimental EL results published in the literature can be described by this model as well.

\section{Experiment}
\begin{figure}[t!] 
\includegraphics[scale=0.3]{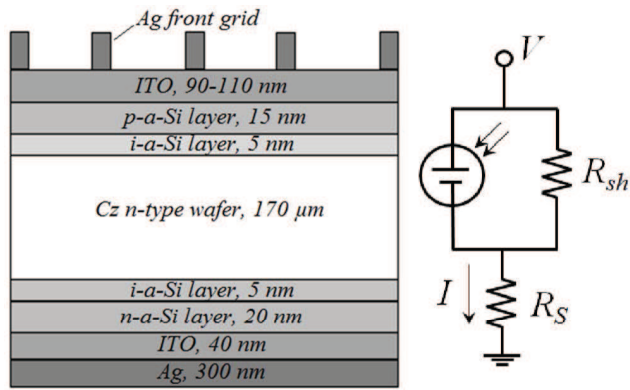}
\caption{Schematic illustration of the HJSC elements used in this study (left), and the equivalent circuit used to derive the $J - V$ relation (right).}
\label{fig1}
\end{figure}

The structure of the HJSCs studied in this work is shown in Fig.~\ref{fig1}. It is based on an n-type c-Si wafer of (100) orientation fabricated using Czochralski's method. The wafers, cut out of the same part of silicon ingot, had the thickness $d = 170\,\mu$m. The HJSC area was 239\,cm$^2$. Deposited on the wafer's front surface are the layers of undoped amorphous silicon (a-Si), amorphous p-type silicon (p-a-Si:H), a conducting transparent indium tin oxide (ITO) layer, and a silver grid for current collection. On the rear side are a layer of a-Si, an n-a-Si:H layer, an ITO layer, and a silver layer. The parameters of the so fabricated HJSCs are given in Table~\ref{table1}, first row. Because all wafers were cut from the same silicon crystal, they were characterized by the same Shockley-Read-Hall lifetime, $\tau_{SRH}$, and bulk recombination velocity $d/\tau_{SRH} = 8.5$\,cm/s. Their radiative recombination velocity at 300 K was not higher than 1 cm/s.

\begin{table*}[t!]
\centering
\caption{Parameters of our HJSC. $J_{SC}$, $V_{OC}$, and $P$ are given for $T = 300$\,K under AM1.5G conditions. In the calculations, the shunt resistance was set to $R_{sh} = 20$\,k$\Omega$, but the results remained practically the same also for larger values of $R_{sh}$. Shockley-Read-Hall recombination lifetime in the space-charge region $\tau_R = 1.2\,\mu$s, and the ratio of hole and electron capture cross section by a deep recombination level $b = 0.01$, see Eq.~(\ref{2}). Bulk Shockley-Read-Hall lifetime $\tau_{SRH}$ was experimentally established by measuring photoconversion decay time. Doping level was measured from Hall effect measurements. Surface recombination velocity, $S$, and series resistance, $R_S$, of our samples were found by fitting the experimental and theoretical open-circuit voltage and photoconversion efficiency, see Eqs.~(\ref{7}) and (\ref{12}) below. In total, we investigated 93 samples, which differed in surface recombination velocity. As a result, $J_{SC}$, $V_{OC}$, and $\eta$ varied among the samples as well. In the table, variation range of these parameters is indicated. For comparison, parameters of the HJSCs from Refs.~\cite{Jano13} and \cite{Green09} are also given in the last two rows. }
\begin{tabular}{| l | l | l | l | l | l | l | l | l | l |}
\hline
Ref. & $N_d$, cm$^{-3}$ & $d$, $\mu$m & $\tau_{SRH}$, ms & $J_{SC}$, mA/cm$^2$ & $V_{OC}$, mV & $\eta$, \% & $S$, cm/s & $A_{SC}$, cm$^2$ & $R_S$, $\Omega$\\ \hline
This work & $10^{15}$ & 170 & 2 & 36-37 & 690-730 & 17.8-20.1 & 10-200 & 239 & 0.0033\\ \hline
\cite{Jano13} & $4.9\cdot 10^{15}$ & 98 & 3.8 & 39.5 & 750 & 24.7 & 1.5 & 100 & 0.0027 \\ \hline
\cite{Green09} & $9.3\cdot 10^{15}$ & 200 & 1 & 42.7 & 706 & 25 & 47 & 4 & 0.13 \\ \hline
\end{tabular}
\label{table1}
\end{table*}

The chemical processing of the samples involved sequential transition of a batch with 24 wafers through vessels with different chemical solutions and included the following stages: cleaning-up the surfaces from contaminations, texturing in a KOH base solution with a surface-active additive (isopropyl alcohol) to form pyramidal morphology, and final cleaning. The subsequent stages consisted of plasma-enhanced chemical vapor deposition of the a-Si layers, deposition of transparent indium tin oxide (ITO) electrodes, and contact grid formation. These technological stages are described in greater detail in Refs.~\cite{Chu09, Zubel12}. 

\begin{figure}[t!] 
\includegraphics[scale=0.75]{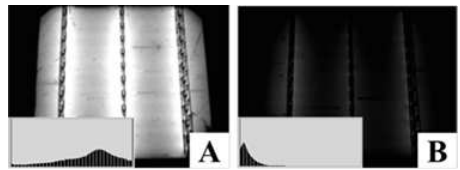}
\caption{Photographs of two representative samples with relatively strong (A) and weak (B) EL intensity.}
\label{fig2}
\end{figure}

The texturing process took place in the same chemical solution of volume 10\,liters, which gradually accumulated the reaction products of KOH with silicon, such as potassium silicate: Si + 2KOH+ H$_2$O = K$_2$SiO$_3$ + 2H$_2\uparrow$. The process proceeded at the temperature between 70 and 80\,$^\circ$C, at which intensive hydrolysis of K$_2$SiO$_3$ takes place, resulting in the formation of silicon dioxide gel. At larger potassium silicate concentrations, the SiO$_2$ film makes transport of the reagents to silicon surface more difficult, leading to a notable dispersion of etching rates among different microscopic regions of silicon surface, and hence to the sample texturing. Apparently, the probability of defect and step formation is higher on the pyramid faces, and those defects act as recombination centers, which have a negative effect on the final EL and photoconversion efficiency.

Even the samples belonging to the same batch were treated under somewhat different chemical conditions due to the unavoidable local inhomogeneity of the solution and the non-equivalence of sample positioning within the batch. Because of this, the samples were characterized by the total (front and rear) surface recombination velocity. By fitting our experimental results with the theory described in the next section, we established that it ranged from 10 to 200\,cm/s. 

The measurements of our solar cells' characteristics at 300 K were performed using cetisPV-Celltest3 setup (HALM, Germany) equipped with a cetisPV-cell-EL-Lab system, a 1.4 MP CCD camera, and light filters that cut off the wavelengths below 880 nm.  For photoconversion measurements, we used a simulator of AM1.5G solar radiation spectrum of 1 kW/m$^2$ power from Yamashita Denso Corp. (Model SS-80AA, class AAA); see \cite{Sachenko16b} for further experimental details. 93 HJSCs were tested. Shown in Fig.~\ref{fig2} are the photographs of two representative samples characterized by the relatively strong and weak EL intensity (the brightness histograms and intensity values were obtained by processing the EL maps in Photoshop).

EL intensity measurements were performed at constant dark current density  $J = 37.65$\,mA/cm$^2$ maintained by an external bias. Note that as surface recombination velocity increases, so does the recombination current through the sample, and thus smaller applied bias is required to maintain this current density. 

\section{Theoretical modeling of EL and photoconversion efficiency of HJSCs}

\subsection{Uniformity of excess carrier concentration profile}
\label{uniformity}
Excess minority hole concentration, $\Delta p(x)$, in the base n-Si region of thickness $d$ is determined from the diffusion equation
\begin{equation}
\frac{d^2\Delta p(x)}{dx^2} - \frac{\Delta p(x)}{L_d} = 0
\label{a1}
\end{equation}
with the solution
\begin{equation}
\Delta p(x) = C_1e^{-x/L_d} + C_2 e^{x/L_d}\ .
\label{a2}
\end{equation}
The constants $C_{1, 2}$ are to be determined from the boundary conditions at $x = 0$ and $x = d$.  

If the diffusion length greatly exceeds the base thickness $d$, $\Delta p(x)$ can be treated as approximately constant. To derive the criterion under which this approximation is permissible, let us expand the two-exponential solution in $x$ to the first order around $x = d/2$:
\begin{equation}
\Delta p(x) = \Delta p^* + C\frac{x - d/2}{L_d} + \ldots 
\label{a3}
\end{equation}
with $\Delta p^* = \Delta p(x = d/2)$ and $C/L_d = d\Delta p/dx|_{x = d/2}$. Obviously, the first-order term can be neglected if the differences $|\Delta p(x = 0) - \Delta p^*|$ and $|\Delta p(x = d) - \Delta p^*|$ are both much smaller than $\Delta p^*$. This gives us the criterion
\begin{equation}
\Delta p^* \gg |C|\frac{d}{2L_d}\ .
\label{a4}
\end{equation}
To establish the relation between $C$ and $\Delta p^*$, we use the boundary condition at $x = d$:
\begin{equation}
S_d\Delta p(d) + D\frac{d\Delta p}{dx}(x = d) = 0\ ,
\label{a5}
\end{equation}
where $S_d$ is surface recombination velocity in the $x = d$ plane. Substitution of the first-order approximation (\ref{a3}) into (\ref{a5}) gives $C = -S_dL_d\Delta p^*/(D + (dS_d/2))$. Combining this expression with the criterion (\ref{a4}), we obtain a simple condition under which the excess concentration $\Delta p(x)$ in the base region can safely be approximated with a constant $\Delta p^*$:
\begin{equation}
S_d \ll 2D/d\ ,\ \ {L_d \gg d/2}\ .
\label{6a}
\end{equation}
 
Hole diffusion coefficient in silicon at $N_d = 10^{15}$\,cm$^{-3}$ is about 10\,cm$^2$/s. Then, for $d = 170\,\mu$m, the theory presented below is valid for $S_d \ll 1200\,\text{cm/s}$. Given that the total (front plus rear) surface recombination velocity in the $x = 0$ and $x = d$ planes was found to be smaller than 200\,cm/s (see below), the {first} criterion (\ref{6a}) is well fulfilled in our experiment. {The second criterion, expressing the possibility of linearizing the expression (\ref{a2}) around $x = d/2$, is likewise fulfilled, because the diffusion length $L_d \approx \sqrt{D\tau_{SRH}} = 1.4$\,mm exceeds sample thickness $d = 170\,\mu$m by an order of magnitude.} In the following, we will omit the asterisk and write $\Delta p$ instead of $\Delta p^*$.

\subsection{Dark $J - V$ curves}
The current through the structure equals to the sum of the current through the shunt resistance, $(V - IR_S)/R_{sh}$, and through the HJSC (see Fig.~\ref{fig1}). We focus on the case specified by the inequality (\ref{6a}), assuming that excess carrier concentration, $\Delta p$, in crystalline Si is practically uniform. The validity of this assumption in our experiment will be justified in the next section. The current through the HJSC is due to the recombination mechanism and is given by $qA_{SC}d \Delta p/\tau_{eff}$, where $q$ is the elementary charge and $\tau_{eff}$ is the effective lifetime in the c-n-Si. Dividing both contributions -- the shunt resistance current and the recombination current in the HJSC -- by the SC area, $A_{SC}$, we obtain the net current density
\begin{equation}
J(V) = \frac{qd\Delta p}{\tau_{eff}} + \frac{VA_{SC}^{-1} - J(V)R_S}{R_{sh}}\ .
\label{1}
\end{equation}
Voltage drop across the HJSC is $V - A_{SC}JR_S$. On the other hand, it equals to the sum of potential differences between the central n-Si slab and the upper and lower amorphous layers of n-a-Si and p-a-Si. These two voltages can be expressed in terms of the equilibrium, $n_0$ and $p_0$, and the excess, $\Delta p$, charge carrier concentrations as
\begin{equation}
V - A_{SC}JR_S = \frac{kT}{q}\left(\ln\frac{n_0 + \Delta p}{n_0} + \ln\frac{p_0 + \Delta p}{p_0}\right)\ .
\label{2a}
\end{equation}
Multiplying both sides by $q/kT$ and performing exponentiation, we obtain a quadratic equation for the excess carrier concentration. The solution of this equation reads
\begin{eqnarray}
&&\Delta p(V) = -\frac{n_0 + p_0}{2} \nonumber \\
&&\ \ + \sqrt{\frac{(n_0+p_0)^2}{4} + n_i^2\left(e^{q(V - JA_{SC}R_S)/kT} - 1\right)} \approx \nonumber\\
&&\ -\frac{N_d}{2} + \sqrt{\frac{N_d^2}{2} + n_i^2\left(e^{q(V - JA_{SC}R_S)/kT} - 1\right)}\ .
\label{5}
\end{eqnarray}
Here, we used the relation $n_0p_0 = n_i^2$, where $n_i$ is intrinsic carrier density. For the experimentally relevant doping levels such that $N_d \gg n_i$, we have set $n_0$ to $N_d$ and neglected $p_0$ in the last line.

The inverse effective lifetime in crystalline Si consists of the bulk and the surface contributions:
\begin{equation}
\tau_{eff}^{-1} = \tau_b^{-1} + (S + U_{SC})/d\ ,
\label{5a}
\end{equation}
where $S$ is the net surface recombination velocity in the $x = 0$ and $x = d$ planes, i.e. on the boundaries between crystalline and amorphous Si layers. The recombination velocity in the space-charge region (SCR) is given by \cite{Shockley57}
\begin{eqnarray}
&&U_{SC}(V) = \frac{1}{\tau_R}\int_0^w dx\,\left(N_d + \Delta p(V)\right)\nonumber \\
&&\ \ \ \ \big[(N_d + \Delta p(V))e^{-y_{pn}(1 - \frac{x}{w})^2} \nonumber \\
&& \ \ \ \ \ \ \ \ \ \ + b\Delta p(V)e^{y_{pn}(1 - \frac{x}{w})^2}\big]^{-1}\ .
\label{2}
\end{eqnarray}
Here, $\tau_R$ is the Shockley-Read-Hall lifetime in the SCR, $b = \sigma_p/\sigma_n$ is the ratio of hole and electron capture cross sections by a deep recombination level,  $w = 2\sqrt{\ln(N_d/n_i)\varepsilon_0\varepsilon_skT/(q^2N_d)}$ is the SCR thickness, $\varepsilon_S = 11.7$ is the relative permittivity of Si, and $y_{pn} = 2\ln(N_d/n_i(T)) - qV/kT$ is the dimensionless inversion potential. The numerical values of the parameters used are given in Table~\ref{table1}. 

The bulk lifetime is determined by several recombination mechanisms:
\begin{equation}
\tau_b = \left(\frac{1}{\tau_{SRH}} + \frac{1}{\tau_r} + \frac{1}{\tau_{nr}} + \frac{1}{\tau_{Auger}}\right)^{-1}\ ,
\label{3}
\end{equation}
where the first term, $\tau_{SRH}$, is the bulk Shockley-Read-Hall lifetime in c-Si, the second and the third terms,
\begin{equation}
\tau_r = \frac{1}{A(n_0 + \Delta p(V))}\ ,\ \ \tau_{nr} = \frac{\tau_{SRH}n_x}{n_0 + \Delta p(V)}
\label{4}
\end{equation}
are, respectively, the radiative recombination lifetime characterized by the parameter $A$ \cite{Sachenko06}, and the exciton non-radiative recombination lifetime \cite{Sachenko16}. The latter depends on the parameter $n_x$, whose value in Si at room temperature is $8.2\cdot10^{15}$ cm$^{-3}$ \cite{Sachenko16}. Finally, $\tau_{Auger}$  is Auger band-to-band recombination lifetime, whose rather involved expression is taken from the work \cite{Richter12} and is not reproduced here.  In our case, Auger recombination mechanism practically does not play a role, but it becomes important at high doping levels \cite{Sachenko15}; therefore, it is included in Eq.~(\ref{3}) for the sake of generality.

\subsection{$J - V$ curves in the presence of illumination}
Illumination of the HJSC results in the onset of the light-generated current in the reverse direction. The expression for the excess carrier concentration (\ref{5}) remains the same, but the current expression becomes:
\begin{equation}
J(V) = \frac{qd\Delta p}{\tau_{eff}} + \frac{V A_{SC}^{-1} - J(V)R_S}{R_{sh}} - \Delta J\ ,
\label{9}
\end{equation}
where the light-generated part of the current density, $\Delta J$, is chosen so as to have the correct short-circuit current density limit, $J(V = 0) = –J_{SC}$:
\begin{equation}
\Delta J = qd\left(\Delta p/\tau_{eff}\right)_{SC} + J_{SC}\left(1 + R_S/R_{sh}\right) \approx J_{SC}\ ,
\label{7a}
\end{equation}
where $(\Delta p/\tau_{eff})_{SC}$ is given by (\ref{5}), (\ref{5a}) with $V = 0$ and $J = -J_{SC}$. The short-circuit current needs to be determined experimentally. Its substitution into (\ref{9}) and numerical solution of this equation allows one to describe the $J - V$ curve at arbitrary bias values. In particular, the open-circuit voltage is [cf. (\ref{2a})]
\begin{equation}
V_{OC} = \frac{kT}{q}\ln\frac{(N_d + \Delta p_{OC})\Delta p_{OC}}{n_i^2(T)}\ .
\label{7}
\end{equation}
The excess electron-hole pair concentration in silicon bulk under the open-circuit conditions, $\Delta p_{OC}$, should be found from the generation-recombination balance equation,
\begin{equation}
J_{SC} = q\left(S + (d/\tau_b)\right)\Delta p_{OC}\ .
\label{8}
\end{equation}

Using the maximal-power condition,  $d(VJ(V))/dV = 0$, the voltage $V_m$ at the maximal power can be found. Substitution of $V_m$ into Eq.~(\ref{9}) allows determining the respective photocurrent $J_m$. As a result, an expression for the maximal photoconversion power is obtained,
\begin{equation}
P = A_{SC}J_mV_m\ .
\label{11}
\end{equation}
The $J - V$ curve fill factor and photoconversion efficiency are
\begin{equation}
FF = \frac{J_mV_m}{J_{SC}V_{OC}}\ ,\ \ \eta = \frac{J_m V_m}{P_S}\ ,
\label{12}
\end{equation}
where $P_S$ is the incident radiation energy flux. 

\subsection{Electroluminescence intensity}
Experimental EL intensity,
\begin{equation}
{I_{EL} = \frac{J}{q}\gamma_{int}\gamma_{ext}\ , }
\label{13}
\end{equation}
{at constant current $J$} is proportional to the internal EL quantum efficiency
\begin{equation}
\gamma_{int} = \tau_{eff}/\tau_r ,
\label{6}
\end{equation}
where all relaxation times are given by (\ref{5a})-(\ref{4}). {The external EL efficiency $\gamma_{ext}$ is of the order of a few per cent and is affected by the total internal reflection by the front surface at large angles of incidence, Fresnel reflection at small angle of incidence, and by optical losses due to photon reabsorption. Note that the effect of the last factor is the smallest of all three, because Because surface recombination velocity is not among the factors that influence $\gamma_{ext}$, we will treat $\gamma_{ext}$ merely as a fit parameter without attempting to determine it theoretically.}

\section{Experimental results and discussion}
\subsection{Surface recombination effect on EL and photoconversion efficiency of HJSCs under AM1.5 conditions}
\begin{figure}[t!] 
\includegraphics[scale=0.3]{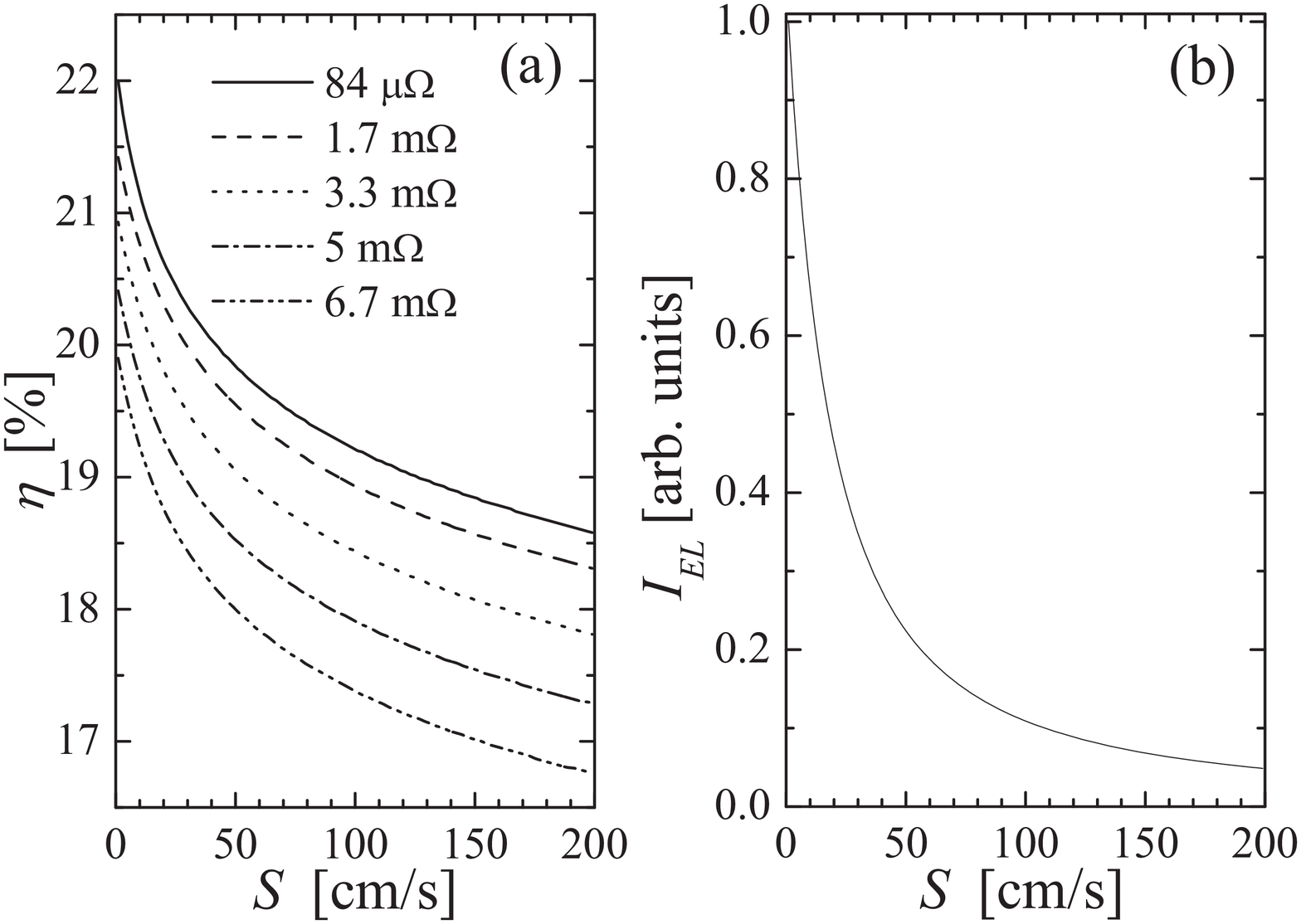}
\includegraphics[scale=0.3]{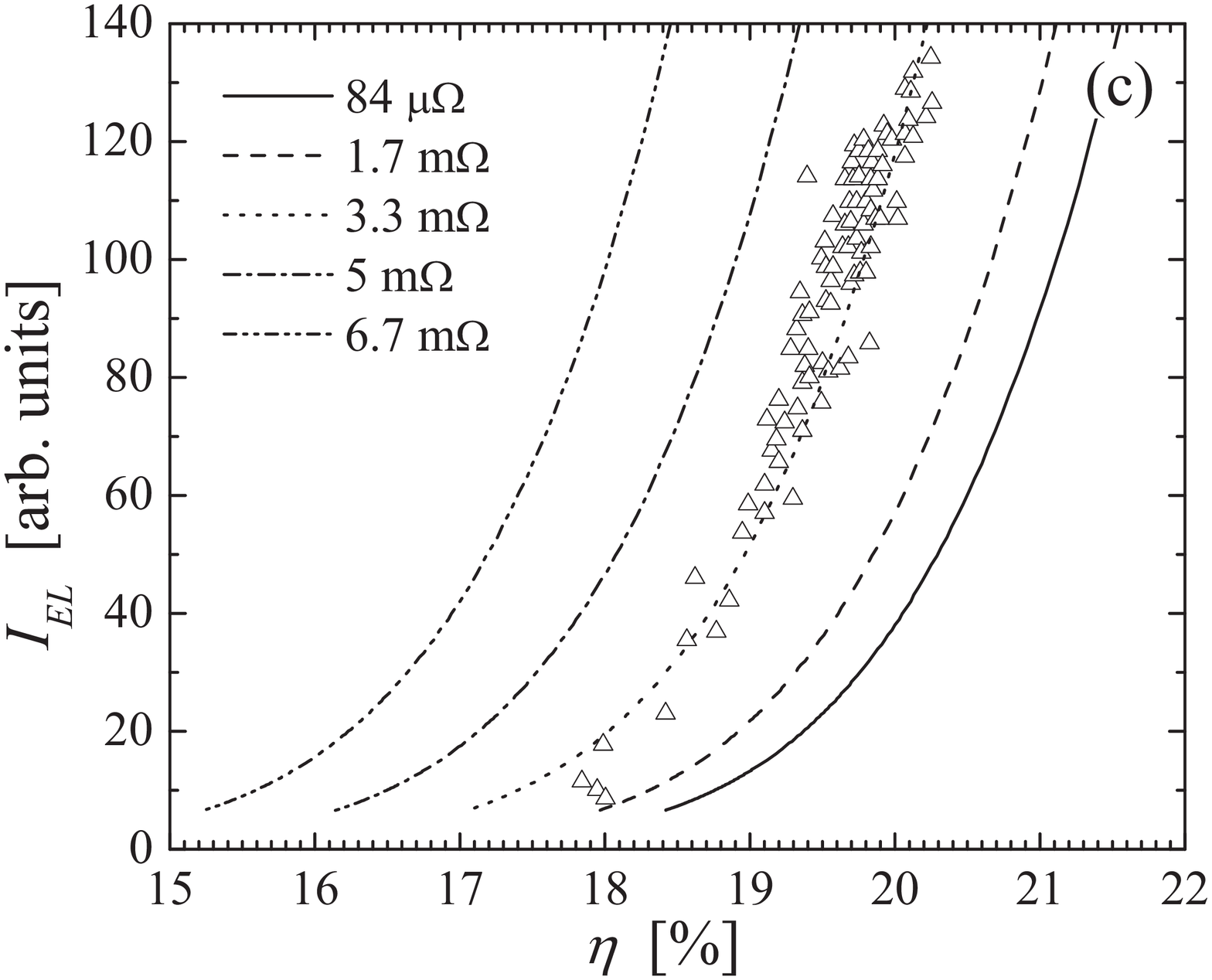}
\caption{Theoretical photoconversion efficiency (a) and EL intensity (b) vs. surface recombination velocity. In panel (a), the curves are obtained for different series resistance $R_S$, as indicated in the legend. In panel (b), the $El$ curves are practically the same for all these values of $R_S$. (c) EL intensity vs. photoconversion efficiency.}
\label{fig5}
\end{figure}

While all samples studied were characterized by very close values of photoconversion efficiency from 17.8\,\% to 20.1\,\%, EL exhibited a large dispersion among the samples. To understand this result, shown in Fig.~\ref{fig5}(a) is the theoretical dependence of photoconversion efficiency on the surface recombination velocity. Different curves in this figure are built for different series resistance values. For comparison, Fig.~\ref{fig5}(b) shows the theoretical EL intensity vs. $S$ at the current density of 37.68\,mA/cm$^2$. As seen from panels (a) and (b) of Fig.~\ref{fig5}, EL intensity decreases with $S$ much faster than photoconversion efficiency does. Indeed, if the latter decreases by about 10\%, the former decreases by more than an order of magnitude. 

\begin{figure}[t!] 
\includegraphics[scale=0.3]{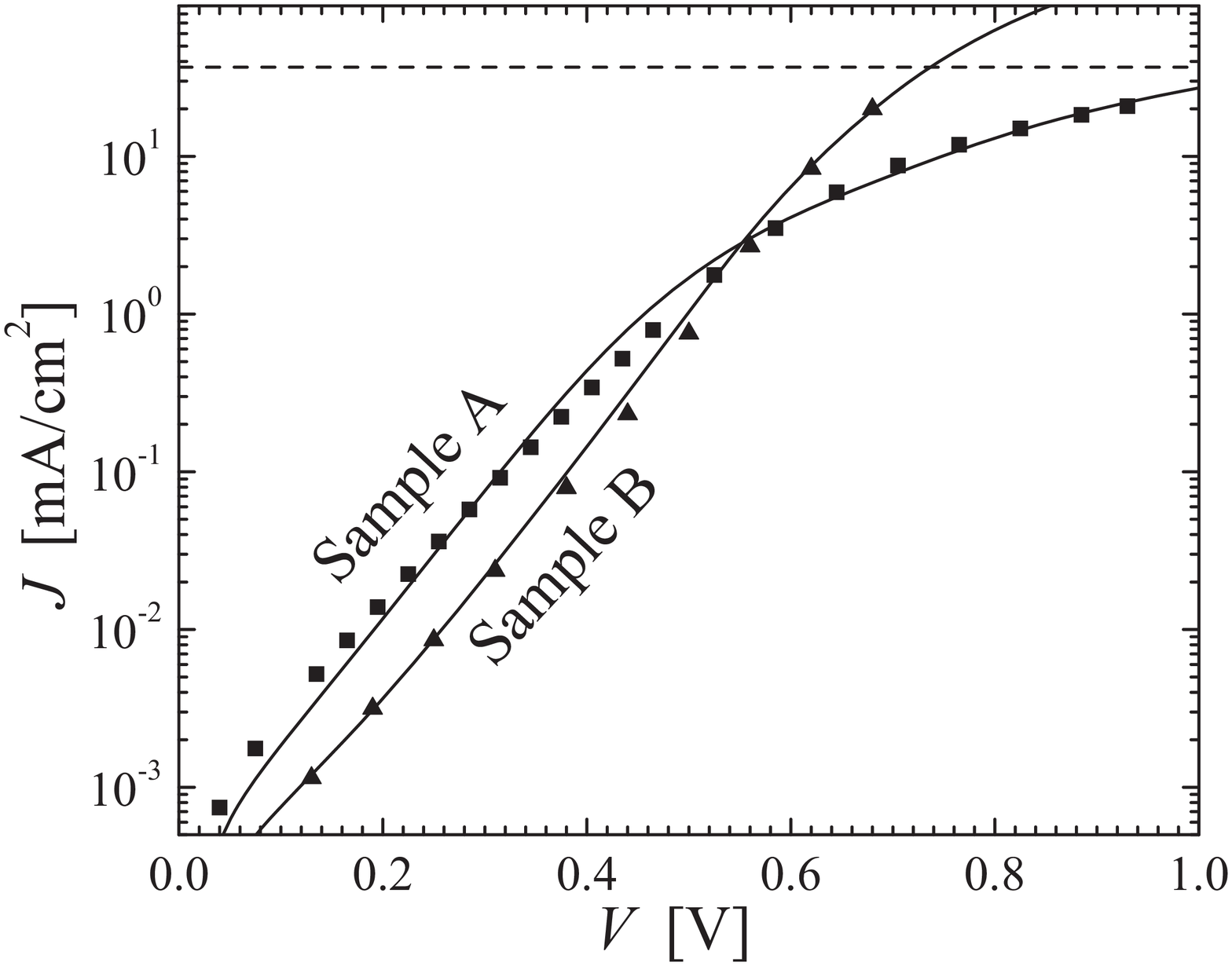}
\caption{Dark $J - V$ curve under AM1.5 illumination conditions for  two small-area samples that differ in series resistance ($R_S = 1.5\,\Omega$ for sample A and $12\,\Omega$ for sample B. Surface recombination velocity $S = 10$\,cm/s for both samples. The theoretical curves are obtained using Eqs. (\ref{1})-(\ref{4}). The horizontal line is drawn at a 36.76 mA/cm$^2$ level.}
\label{fig3}
\end{figure}

\begin{figure}[t!] 
\includegraphics[scale=0.3]{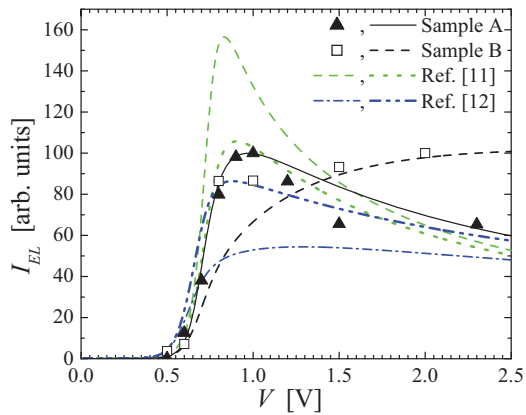}
\caption{Experimental integral EL intensity vs. applied voltage for the two samples (symbols), and the theoretical fit curves (black solid and dashed lines). The theoretical curve for sample A is obtained for the series resistance $R_S = 1.5\,\Omega$, and for sample B, $R_S = 12\,\Omega$. The shunt resistance was $R_{sh} = 10$\,k$\Omega$ and the surface recombination velocity $S = 10$\,cm/s for both samples. The green curves were built for parameter values from Ref.~\cite{Jano13}, but with $S = 1.5$\,cm/s (dashed line) and $S = 12$\,cm/s (dotted line). The blue curves were built for the parameters taken from Ref.~\cite{Green09}, but with $S = 47$\,cm/s (dashed) and $S = 1.5$\,cm/s (dotted).}
\label{fig4}
\end{figure}

\begin{figure}[t!] 
\includegraphics[scale=0.3]{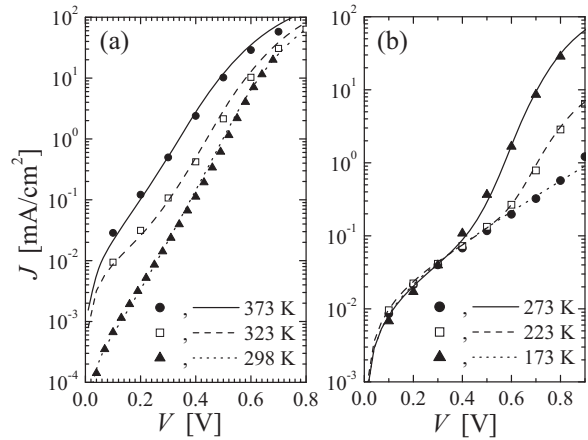}
\caption{Dark $J - V$ curves for sample A at different temperatures. The theoretical curves are built using Eqs. (\ref{1})-(\ref{4}).}
\label{fig7}
\end{figure}

Fig.~\ref{fig5} (c) shows the experimental (symbols) and theoretical (lines) photoluminescence intensity $El$ vs. photoconversion efficiency taken for all samples studied (symbols) and the theoretical curves obtained for different series resistance values. The fact that all experimental data points fit a single master curve predicted by the theory confirms its validity. Comparison between the two sets of data allows determining the series resistance to be $R_s = 3.3$\,m$\Omega$. Besides, the value of surface recombination velocity can be determined from this plot for each individual sample. The larges $S$-value so found does not exceed 200\,cm/s. While it is impossible to distinguish between front and rear surface contributions to $S = S_0 + S_d$, it is reasonable to assume that they are about the same. This means that the rear surface recombination velocities in our samples $S_d < 100$\,cm/s, as required by the criterion (\ref{6a}).

These results indicate that EL intensity of HJSCs can serve as an indicator of their semiconductor structure quality, in particular, of the interface between the textured wafer and the heterojunction. 

For temperature-dependent measurements (see next subsection), one of the large-area HJSC samples was cut into small pieces with the area of $A_{SC} = 1$\,cm$^2$. Each of those small-area samples was treated individually to produce the front and rear metal contacts, and therefore was characterized by its individual parasitic resistance. 

Shown in Fig.~\ref{fig3} are the experimental dark $J - V$ curves for two representative small-area HJSC samples, measured at $T$ = 300 K. The two samples were cut off from the same large-area SC characterized by the surface recombination velocity $S = 10$\,cm/s; their photoconversion parameters $J_{SC}$, $V_{OC}$, and $\eta$ are indicated as upper limits of the respective ranges in Table~\ref{table1}, i.e. 37\,mA/cm$^2$, 730\,mV, and 20.1\,\%, respectively. The respective theoretical curves agree with the experimental data if one takes the series resistance of samples A and B to be 1.5\,$\Omega$ and 12\,$\Omega$, respectively.

Fig.~\ref{fig4} shows the experimental integral EL intensity vs. direct bias for these two samples. To obtain the experimental $El$, the measured EL spectrum was integrated from $1.02\, \mu$m to $1.28\,\mu$m [see (\ref{13})], as $I_{el}(\lambda)$ was practically zero outside of these limits. The theoretical internal EL quantum efficiency $\gamma_{int}(V)$ was found using Eq.~(\ref{6}). The agreement between the theoretical and experimental curves can be achieved only by taking into account the dependence of the radiative recombination coefficient in silicon on the injection level $\Delta p(V)$, see Ref.~\cite{Sachenko06}. 

For sample A, this curve is non-monotonic, developing a maximum at around 1 V, whereas for sample B, it monotonically increases, similar to the respective curves from Ref.~\cite{Ng01}. All theoretical curves in Fig.~\ref{fig4} were normalized to the maximal $\gamma_{int}$ value of the sample A and post-multiplied by 100; the respective experimental data points were similarly normalized to the maximal EL intensity for sample A. The maximal theoretical value of the internal EL quantum efficiency was 2.6\%, which is somewhat larger than the experimental EL external efficiency -- the ratio of the irradiated to absorbed power -- of 2.1\%. Although the internal quantum efficiency of EL is greater than the external quantum efficiency, in the case of HJSCs studied here, the difference between them is quite small. The first reason for this is that the part of the surface area shadowed by the contacts is only a few per cent of the total area. Second, both sides of the silicon wafer are microtextured, which leads to a strong reduction of light reflection coefficient from the surface.

The dashed green curve in Fig.~\ref{fig4} is the theoretical $\gamma_{int}(V)$ dependence for HJSC parameters from Ref.~\cite{Jano13} (see the second row of Table~\ref{table1}). Note that for the samples studied in Ref.~\cite{Jano13}, surface recombination velocity is quite low (about 1.5\,cm/s). Recalculation of the maximal quantum efficiency for the case when $R_S = 1.5\,\Omega$ and $V = 0.83$\,V gives the value of 4.1\%.

The dotted green curve in Fig.~\ref{fig4} represents the theoretical $\gamma_{int}(V)$ dependence for HJSC parameters from the same Ref.~\cite{Jano13}, except for the surface recombination velocity, $S$, which was set to 12\,cm/s. As seen from Fig.~\ref{fig4}, the difference between the peak EL efficiency obtained for this set of parameters and for the fitting curve of sample A is about 6\%, whereas the peak values for the green dashed and dotted curves differ by 57\%. The main reason for this discrepancy is low surface recombination velocity value realized in Ref.~\cite{Jano13}, whereas other characteristics, such as wafer thickness, doping level, Shockley-Read-Hall lifetime, and short-circuit current affect EL quantum efficiency much more weakly.

Finally, the dash-dotted blue curve in Fig.~\ref{fig4} shows the theoretical $\gamma_{int}(V)$ dependence for a p-n junction-based SC with photoconversion efficiency of 25\% \cite{Green09}. The parameters used to build this curve are given in the third row of Table~\ref{table1}. Surface recombination velocity of 47\,cm/s was determined in the work \cite{Sachenko15} by comparing the main theoretical and experimental characteristics of SCs. As seen from comparison of this curve with the results obtained for Sample A, the theoretical EL internal quantum efficiency value is about twice as small as in the HJSC element studied here (ca. 1.3\%), which is quite close to the experimental external quantum efficiency from Ref. ~\cite{Green01}. Setting $S$ to 1.5\,cm/s, the peak EL quantum efficiency of 2.25\% is obtained. Photoconversion efficiency of both HJSCs and p-n junction-based solar cells is a much weaker function of the surface recombination velocity than EL efficiency. As the $\eta$-values from Table~\ref{table1} suggest, in spite of the rather large $S$ in p-n junction-based solar cells, the photoconversion efficiency in them is actually the highest.

It should be noted that the theoretical value of the EL internal quantum efficiency, $\gamma_{int}$, is strongly influenced by the non-radiative exciton recombination. For example, for the first sample, neglecting this effect leads to an increase of $\gamma_{int}$ at maximum to 4.08\%, as compared to the value of 2.6\% obtained with taking this recombination channel into account, i.e. the difference is by a factor 1.5. At the same time, photoconversion efficiency at the parameter values considered here is influenced by this mechanism by only about 2\%, see Ref.~\cite{Sachenko16}. This is related to the difference in the effect of the total surface plus bulk recombination velocity on the EL and photoconversion efficiency, which can be traced back to the fact that EL internal quantum efficiency is inversely proportional to the recombination current, whereas photoconversion efficiency is inversely proportional to the logarithm of recombination current. Thus, the general conclusion is that the maximal EL efficiency both in HJSCs and in p-n junction-based SCs is realized for the minimal total (surface + bulk) recombination velocity.

\subsection{Temperature dependence of electroluminescence in HJSCs}
Shown in Fig.~\ref{fig7} are the experimental $J - V$ curves for sample A, measured at the temperatures $T$ = 173, 223, 273, 298, 323, and 373 K. They correlate well with the dark and illuminated $J - V$ curves from Ref.~\cite{Sachenko16b}. As seen in the figure, theoretical curves are in good agreement with the experimental ones. An interesting feature of the experimental $J(V)$ curves for the temperatures 173, 223, and 273 K, see Fig.~\ref{fig7}(b), is that at the current densities below 1\,mA/cm$^2$, the current is purely due to tunneling mechanism. This is seen from the current reduction as compared to the current in Si-based p-n junctions (see, e.g., \cite{Ng01}), as well as from the strong decrease of the slope of the $J - V$ curves. The expression for the tunneling current
\begin{equation}
J_T = J_0\left(e^{qV/E_T} - 1\right)\ ,
\label{14}
\end{equation}
where $J_0$ is saturation current density and $E_T$ tunneling energy, together with the expressions (\ref{1})-(\ref{4}) at $J > 1$\,mA/cm$^2$, allows to obtain good agreement of the theory with the experiment if one takes $J_0 = 1.4\cdot10^{-2}$\,mA/cm$^2$ and $E_T$ = 0.21 eV. 

In Ref.~\cite{Sachenko16b} it was shown that the series resistance increases on cooling due to the increase of the resistance in the ITO-metal contact. This observation correlates well with the results of Ref.~\cite{Sachenko17}.

\begin{figure}[t!] 
\includegraphics[scale=0.3]{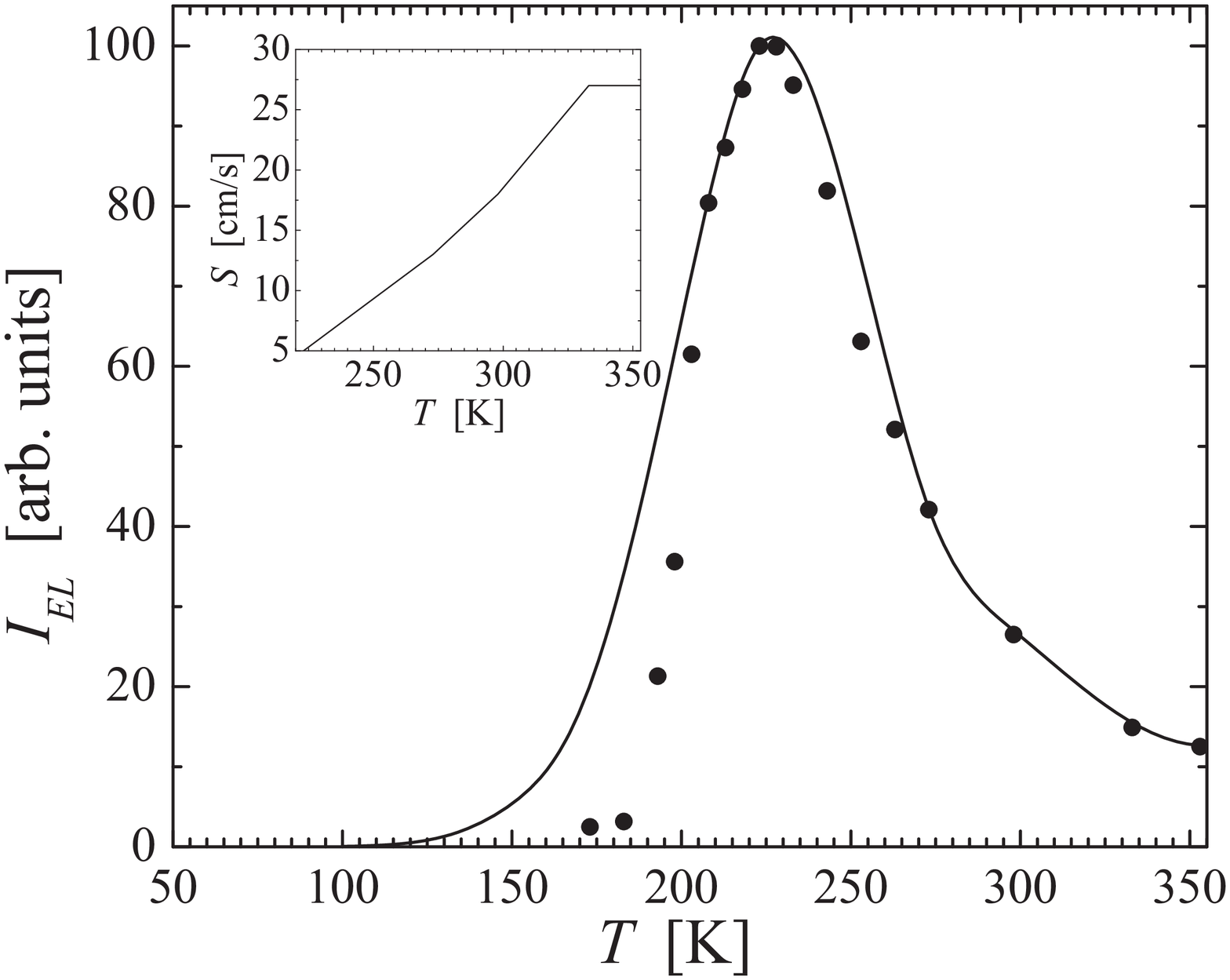}
\caption{Experimental (symbols) and theoretical (solid curve) EL intensity vs. temperature. The theoretical curve is built using Eqs.~(\ref{6}) and (\ref{2})-(\ref{4}). The inset shows the assumed temperature dependence of surface recombination velocity.}
\label{fig8}
\end{figure}

Fig.~\ref{fig8} shows the experimental temperature dependence of the integral EL intensity measured at the recombination current density of 37.65~mA/cm$^2$, normalized to the maximal value at $T = 223$\,K. At higher temperatures, EL intensity decreases due to the decrease of the radiative recombination coefficient with temperature. At low $T < 200$\,K, radiation recombination parameter $A$ is determined mostly by exciton radiative recombination, and can be written as
\begin{equation}
A \approx \frac{N_x}{N_cN_v}\frac{1}{\tau_{rx}}e^{E_x/kT}\ ,
\label{15}
\end{equation}
where $N_c$, $N_v$, and $N_x$ are effective densities of states in the conduction, valence, and exciton bands of silicon, $E_x$ is the exciton ground state energy (0.0147\,eV in Si), and $\tau_{rx}$ is exciton radiative recombination lifetime \cite{Sachenko06}. The EL intensity in n-type silicon is proportional to $A(n_0 + \Delta p(V))$. According to Eq.~(\ref{15}), the radiative coefficient increases monotonically on cooling, and therefore its temperature dependence cannot be responsible for the experimentally observed decrease of EL and photoluminescence to the left of the peak in Fig.~\ref{fig8} (see Ref.~\cite{Trupke03}). To explain this effect, one needs to account for the electron freeze-out at low temperatures \cite{Hess}. Equilibrium electron concentration can be found from the neutrality equation
\begin{eqnarray}
&&n_0(T) = \frac{2}{\sqrt{\pi}}N_c(T)\int_0^\infty dx\frac{\sqrt{x}}{1 + e^{x + \frac{E_c - E_F}{kT}}}\nonumber \\
&&\ \ \ \ \ \ \ \ \approx N_ce^{(E_F-E_c)/kT} = \frac{N_d}{1 + 2e^{(E_F - E_d)/kT}}\ ,
\label{16}
\end{eqnarray}
where $E_F$ is Fermi energy, $E_d$ is the energy of donor level, and $N_c(T) = N_{c}(300\,\text{K})(T/300\,\text{K})^{3/2}$ the effective density of states in silicon conduction band at 300\,K. It follows from Eq.~(\ref{16}) that the lower the temperature, the fewer donors are ionized, which leads to a dramatic decrease of $n_0$ on cooling. Solution of this equation for $e^{(E_F - E_c)/kT}$ and back-substitution of this result into the expression for the electron concentration gives at low temperatures $n_0(T) \approx \sqrt{N_cN_d/2}e^{-(E_c - E_d)/(2kT)}$, i.e. a dramatic decrease of the conduction electron concentration on cooling.

Apart from the experimental curve, Fig.~\ref{fig8} also shows the normalized theoretical internal quantum efficiency of EL vs. temperature obtained with the help of Eq.~(\ref{15}), and also taking into account the electron freeze-out in silicon. In addition to exciton radiative recombination, the band-to-band radiative recombination was also accounted for using the expressions from Ref.~\cite{Sachenko06}. It was assumed that $E_d = 0.044$\,eV (corresponding to phosphorus level in Si), and the parameters $n_x$, $S$, $\tau_{SRH}$, and $\tau_{Auger}$ had the same values as at $T$ = 300\,K. As seen in the figure, the locations of the theoretical and experimental EL intensity maxima agree well with each other. 

In order to achieve a good agreement between theory and experiment, temperature dependence of surface recombination velocity, $S$, had to be assumed, see inset in Fig.~\ref{fig8}. Its increase with temperature is due to the shift of Fermi level toward the middle of the bandgap, leading to a variation in the population of surface states, and a corresponding variation in $S$. Therefore, another reason of $El$ decrease at $T > 223$\,K is the growth of surface recombination velocity with temperature.

\begin{figure}[t!] 
\includegraphics[scale=0.3]{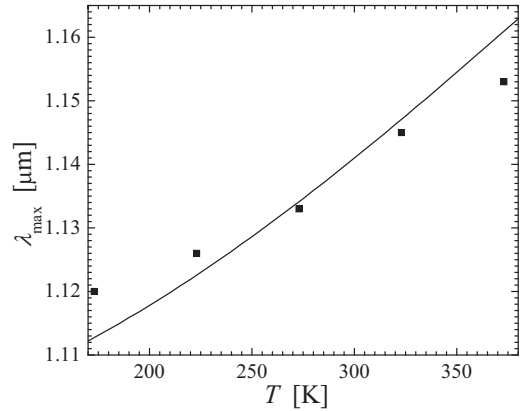}
\caption{Experimental temperature dependence of the maximal EL wavelength (symbols). The theoretical curve is obtained using Eq.~(\ref{17}).}
\label{fig9}
\end{figure}

The relation between the EL intensity reduction in HJSCs to the current transport peculiarities at low temperatures (below 173\,K) is confirmed by the temperature dependence of photoluminescence intensity in silicon from Ref.~\cite{Trupke03}. Photoluminescence intensity also decreases at very low temperatures, but its decrease begins at 125\,K, i.e. 100 degrees below the respective temperature in HJSCs.

Because the binding energy of an exciton in silicon is about 14.7\,meV, EL effect has edge character, and the temperature dependence of the wavelength at which EL has maximal intensity is determined mainly by the temperature dependence of bandgap in silicon (see Fig.~\ref{fig9}). This maximal wavelength is described well by an expression
\begin{equation}
\lambda_{max}(T) = \frac{1240\,\text{nm}}{E_g(T)/1\,\text{eV}} + 38\,\text{nm}\ ,
\label{17}
\end{equation}
where $E_g(T)$ is the bandgap energy. In crystalline silicon, radiative edge luminescence is accompanied by a TO phonon of energy 58\,meV, which alone cannot explain the value of 38\,nm in this expression. It can be assumed that several different radiative recombination processes involving different phonon types may be operative in our HJSCs. 

\section{Conclusions}
{A theoretical model allowing one to quantitatively describe experimental photoconversion efficiency and EL intensity in HJSCs is proposed. Our approximate treatment is valid at the minority diffusion length much greater than sample width and small surface recombination velocities, see Eq.~(\ref{6a}). These conditions are well satisfied in high-efficiency HJSC, but may break down in the standard SCs.}

It is established that the sensitivity of EL in HJSCs and silicon p-n junctions to surface recombination velocity is much stronger than the respective sensitivity of photoconversion efficiency: {EL is inversely proportional to $S$, whereas $\eta$ decreases with $S$ logarithmically}. In HJSCs studied, EL efficiency reaches the value of about 2\%, whereas in silicon p-n junction diodes, it is about 1\%. At the same time, in order to achieve maximal EL efficiency, it is necessary to minimize both surface and bulk recombination velocity.

For correct modeling of EL in HJSCs, it is necessary to take into account the dependence of radiative recombination coefficient on the injection level.

The non-radiative exciton recombination in monocrystalline silicon strongly influences EL internal quantum efficiency in HJSC, in contrast to photoconversion efficiency.

The dark $J - V$ curves of HJSCs were studied in a broad temperature range. Above 298\,K, the net current is formed by the sum of the current through the shunt resistance, recombination current in the space-charge region, and diffusive current, in which the rear surface must be taken into account. At temperatures below 273\,K, in addition to these mechanisms, tunneling current becomes essential.
	
EL intensity is a nonmonotonic function of temperature, having a peak at 223\,K. Its decrease at temperatures below 220\,K can be explained by electron freeze-out at low temperatures. At $T > 220$\,K, it decreases due to the decrease of the radiative recombination coefficient and increase of surface recombination velocity in silicon. 
	
The results obtained here can be used as a methodological basis for surface characterization of monocrystalline silicon.

\section*{Acknowledgments}
M.E. is grateful to the Natural Sciences and Engineering Research Council of Canada (NSERC) and to the Research and Development Corporation of Newfoundland and Labrador (RDC) for financial support.

\end{document}